\documentclass[prb,twocolumn]{revtex4}
\usepackage{amssymb}
\usepackage{makeidx}
\usepackage{amsmath}
\usepackage{graphicx}
\usepackage{dcolumn}
\usepackage{bm}
\usepackage{epstopdf}
\usepackage{color}
\usepackage{txfonts}
\usepackage{microtype}

\begin{document}

\title{High laser harmonics induced by the Berry curvature in time-reversal
invariant materials}
\author{H. K. Avetissian}
\author{G. F. Mkrtchian}
\email{mkrtchian@ysu.am}
\affiliation{Centre of Strong Fields Physics, Yerevan State University, 0025, Yerevan,
Armenia}

\begin{abstract}
A new nonlinear scheme of high harmonics generation in a wide class of
time-reversal invariant materials with broken spatial inversion symmetry
(where recently the nonlinear Hall effect has been established) due to the
nontrivial topology of bands is proposed. A microscopic quasiclassical
theory describing the nonperturbative optical response of
pseudo-relativistic electrons with nonzero Berry curvature of bands to a
strong laser field is developed. We analyze the harmonic content of the
induced current and show that one can decouple induced laser harmonics
solely by the Berry curvature of bands. We also study the dependence of the
nonlinear response on the driving wave and system parameters.
\end{abstract}

\date{\today }
\maketitle

\section{Introduction}

According to Landau's Fermi-liquid theory,\cite{Landau9} the charge
transport in metals involves only quasiparticles with energies near the
Fermi level and depends on bandstructure property at the Fermi level. This
is true for bands with trivial topology. However, for the bands with
nontrivial topology, there is an \textquotedblleft anomalous
velocity\textquotedblright \cite{Kar-Lut}\ of Bloch electrons\ which can be
represented in terms of the Berry curvature\cite{Berry,Xiao2010} of occupied
electronic Bloch states.\cite{Sundaram,Jungwirth,Haldane} At that, the Berry
curvature acts as an effective magnetic field in momentum space.

The Berry curvature being a local gauge field of topological nature depends
on the space-time symmetries of the material. It vanishes in materials that
are symmetric with respect to both spatial and time inversions. In materials
with broken time-reversal symmetry, the Berry curvature leads to an
anomalous Hall effect.\cite{AHE,Qian} There is a wide class of materials
with non-zero Berry curvature in which the spatial inversion symmetry is
broken but the time-reversal is preserved. These are topological insulators,%
\cite{Hasan} transition metal dichalcogenides,\cite{Xiao,Mak,Lensky} gapped%
\textrm{\ }monolayer and bilayer graphene systems,\cite{Gorbachev,Sui} as
well as band-gap modified black phosphorus.\cite{Li,Rodin,Low} In
graphenelike systems the Dirac cones always appear in pairs ($K$, $-K$),\cite%
{Witten} and if the spatial inversion symmetry is broken the Dirac cones
become massive acquiring nonzero Berry curvatures of opposite signs. The
latter is the result of time-reversal symmetry. In contrast to graphenelike
systems in monolayer black phosphorus the conduction and valence band edges
are located at the $\Gamma $ point of the rectangular Brillouin zone.\cite%
{Low}

In the linear response regime, the net topological current identically
vanishes because of time-reversal symmetry.\cite{Ashcroft} Recently, it has
been shown that in the nonlinear response regime, the topological current is
not subject to such symmetry constraints.\cite{Sod-Fu} Two recent
experiments have independently observed second order nonlinear Hall effect
in bilayer\cite{Ma} and multi-layer WTe$_{2}$.\cite{Kang}

The mentioned novel materials with broken inversion symmetry are intensively
considered as an effective medium for the high-order wave mixing and high
harmonic generation (HHG). In particular, HHG has been considered in gapped
graphene\cite{gg} and bilayer graphene systems,\cite{bgg,Ikeda} in
monolayers of black phosphorus,\cite{bf} transition metal dichalcogenides,%
\cite{TMD1,TMD2,2020} hexagonal boron nitride,\cite{HBN} and in buckled
two-dimensional hexagonal nanostructures.\cite{2019a,2019b}

In the last few years, there is a growing interest to use the electronic
response of materials to strong fields for retrieving the electronic
properties of novel nanostructures.\cite{Kruchinin,Schotz} At that the HHG
being the hallmark of strong-field physics plays the central role. The HHG
in solids originates either from the intraband electronic current or from
the interband transitions. In both processes, the topology of bands has a
considerable impact. In particular, Berry curvature affects the excitonic
spectrum in transition metal dichalcogenides along with dynamic energy
modulation due to the Berry connection of bands and consequently has a
sizeable impact on the nonlinear response.\cite{2020,mks2019} The
field-driven injection of electrons across the bandgap strongly depends on
the Berry curvature which defines the structure and timing of HHG spectra.%
\cite{Silva} The topology of bands can also result in distinct by orders
strong-field HHG spectra.\cite{Bauer,Drueke} More interestingly, the
polarization resolved HHG spectra allowed to measure the Berry curvature of $%
\alpha $-quartz.\cite{Luu}

When both interband and intraband mechanisms act simultaneously the sole
contribution of Berry curvature in the nonlinear response is difficult to
separate in materials with time-reversal symmetry. In the highly doped
systems, one can exclude interband transitions and also effectively screen
the many-body Coulomb effects opening the way for the manifestation of
nonlinear topological effects. In this case, only states close to the Fermi
surface will contribute to HHG processes in the low-temperature limit, so
that this response will be a Fermi liquid property as in case of
second-order nonlinear Hall effect.\cite{Sod-Fu} Hence, it is of actual
interest to study the mentioned graphenelike nanostructures physics in the
presence of intense optical fields that can lead to the effective generation
of high harmonics by the Berry curvature of bands.

In the present work, we develop a quasiclassical theory describing the
nonperturbative optical response of pseudo-relativistic electrons with
nonzero Berry curvature of bands to a strong laser field. As a model
time-reversal invariant system we consider massive Dirac nanostructure where
the Dirac cones are tilted.

The paper is organized as follows. In Sec. II the theoretical model and
near-analytical expression for the topological current including
contributions from all orders in the field are presented. In Sec. III, we
examine the harmonic content of the induced current depending on the system
and pump wave parameters. Finally, conclusions are given in Sec. IV.

\section{THEORETICAL MODEL \ \ }

The spectacular transport properties of Dirac materials are connected with
the spinor nature of their electronic wavefunctions and linear dispersion
law around the Dirac points. As was mentioned, the Dirac cones (in
graphenelike systems) always appear in pairs ($K$, $-K$). The nonlinear
topological current is characterized by the odd order moments of Berry
curvatures over occupied states. The second-order nonlinear topological
current is characterized by the first order moment of Berry curvature,
called the Berry curvature dipole. The latter is proposed to exist in
transition metal dichalcogenides,\cite{You,Zhang,Zhou} time-reversal
symmetric Weyl semimetals,\cite{Sod-Fu} in the giant Rashba material BiTeI,%
\cite{Facio} as well as in the strained graphene.\cite{Battilomo} In
addition to inversion symmetry breaking the common feature of all these
materials is the breaking of the three-fold ($C_{3}$) symmetry. The latter
results either low-energy Dirac quasi-particles forming tilted Dirac cones
or trigonal warping of the Fermi surface. In the present paper we will
consider strained nanostructure with tilted Dirac cones. The application of
strain to the lattice deforms the corresponding Brillouin zone, and the
Fermi velocity also becomes anisotropic. The low energy model Hamiltonian up
to the first order in quasimomentum $p$ (relative to $\tau K$ points) will
be:

\begin{equation}
\widehat{H}_{0}=%
\begin{bmatrix}
\Delta +\tau \alpha p_{x} & \tau \mathrm{v}_{x}p_{x}-i\mathrm{v}_{y}p_{y} \\ 
\tau \mathrm{v}_{x}p_{x}+i\mathrm{v}_{y}p_{y} & -\Delta +\tau \alpha p_{x}%
\end{bmatrix}%
,  \label{H111}
\end{equation}%
where $\tau =\pm 1$ is the valley index, $\mathrm{v}_{x}$ and $\mathrm{v}%
_{y} $ are Fermi velocities, $2\Delta $ is the gap. The term $\alpha $
produces a finite tilt in the Dirac cones. This tilting effect is allowed by
symmetry and it is crucial to get a nonvanishing net topological current.
The eigenenergies of the Hamiltonian (\ref{H111}) for valence ($v$) and
conduction ($c$) bands are $\mathcal{E}_{v\tau }\left( \mathbf{p}\right)
=\alpha \tau p_{x}-\mathcal{E}$ and $\mathcal{E}_{c\tau }\left( \mathbf{p}%
\right) =\alpha \tau p_{x}+\mathcal{E}$, with $\mathcal{E=}\sqrt{\Delta ^{2}+%
\mathrm{v}_{x}^{2}p_{x}^{2}+\mathrm{v}_{y}^{2}p_{y}^{2}}$.

We consider the interaction of a strong wave field $\mathbf{E}(t)$ with the
nanostructure. The wave propagates in a perpendicular direction to the
nanostructure plane ($XY$):%
\begin{equation}
\mathbf{E}\left( t\right) =f\left( t\right) E_{0}\hat{\mathbf{e}}\cos \omega
t,  \label{field}
\end{equation}%
with the frequency $\omega $, polarization $\hat{\mathbf{e}}$ unit vector,
pulse duration $\mathcal{T}=40\pi /\omega $, and envelope $f\left( t\right)
=\sin ^{2}\left( \pi t/\mathcal{T}\right) $.

We will consider a highly electron-doped system and we will neglect
excitonic effects since free carriers introduced through doping will
effectively screen the Coulomb interaction. Thus, the dynamics of charge
carriers is described by a single-particle density matrix. We use the second
quantization formalism, expanding the fermionic field operators on the basis
eigenstates of a single particle Hamiltonian (\ref{H111}): $\hat{\Psi}(%
\mathbf{r})=\sum_{\lambda ,\mathbf{k}}\hat{e}_{\lambda \mathbf{k}}|\lambda ;%
\mathbf{k}\rangle e^{i\mathbf{kr}},$ where $\hat{e}_{\lambda \mathbf{k}}$ ($%
\hat{e}_{\lambda \mathbf{k}}^{\dagger }$) are the annihilation (creation)
operators for an electron with momentum $\mathbf{k}$, and $\lambda =\left\{
b,\tau \right\} $ the set of quantum numbers (band and valley). The total
Hamiltonian in the second quantization reads: 
\begin{equation}
\hat{H}=\hat{H}_{\mathrm{free}}+\hat{H}_{\mathrm{int}},  \label{H1}
\end{equation}%
where 
\begin{equation}
\hat{H}_{\mathrm{\ free}}=\sum_{\tau ,\mathbf{p}}\left( \mathcal{E}_{c\tau
}\left( \mathbf{p}\right) \hat{e}_{c\tau \mathbf{p}}^{\dagger }\hat{e}%
_{c\tau \mathbf{p}}+\mathcal{E}_{v\tau }\left( \mathbf{p}\right) \hat{e}%
_{v\tau \mathbf{p}}^{\dagger }\hat{e}_{v\tau \mathbf{p}}\right) 
\label{Hfree}
\end{equation}%
is the free particle Hamiltonian, and 
\begin{equation}
\hat{H}_{\mathrm{int}}=e\mathbf{E}(t)\cdot \widehat{\mathbf{r}}  \label{Hint}
\end{equation}%
is the light-matter interaction Hamiltonian, with the elementary charge $e$
and position operator in the second quantization: $\widehat{\mathbf{r}}=%
\widehat{\mathbf{r}}_{i}+\widehat{\mathbf{r}}_{\mathrm{B}}+\widehat{\mathbf{r%
}}_{e}$. Here 
\begin{equation}
\widehat{\mathbf{r}}_{i}=i\hbar \sum\limits_{\tau ,\mathbf{p,p}^{\prime
}}\delta _{\mathbf{p}^{\prime }\mathbf{p}}\partial _{\mathbf{p}^{\prime
}}\left( \hat{e}_{c\tau \mathbf{p}}^{\dagger }\hat{e}_{c\tau \mathbf{p}%
^{\prime }}+\hat{e}_{v\tau \mathbf{p}}^{\dagger }\hat{e}_{v\tau \mathbf{p}%
^{\prime }}\right)   \label{ri}
\end{equation}%
is the intraband part of position operator and 
\begin{equation}
\widehat{\mathbf{r}}_{\mathrm{B}}=\sum\limits_{\tau ,\mathbf{p}}\mathbf{A}%
_{\tau }\left( \mathbf{p}\right) \left( \widehat{e}_{c\tau \mathbf{p}}^{+}%
\widehat{e}_{c\tau \mathbf{p}}-\widehat{e}_{v\tau \mathbf{p}}^{+}\widehat{e}%
_{v\tau \mathbf{p}}\right)   \label{rb}
\end{equation}%
is defined by the topology of bands with the Berry connection: 
\begin{equation}
\mathbf{A}_{\tau }\left( \mathbf{p}\right) =\hbar \langle b\tau \mathbf{p}%
|i\partial _{\mathbf{p}}|b\tau \mathbf{p}\rangle .  \label{Bcon}
\end{equation}%
The last term 
\begin{equation*}
\widehat{\mathbf{r}}_{e}=\sum\limits_{\tau ,\mathbf{p}}\left( \mathbf{D}_{%
\mathrm{tr}}\left( \tau ;\mathbf{p}\right) \widehat{e}_{v\tau \mathbf{p}}^{+}%
\widehat{e}_{c\tau \mathbf{p}}+\mathrm{h.c.}\right) 
\end{equation*}%
defines interband current with the transition dipole moment $\mathbf{D}_{%
\mathrm{tr}}\left( \tau ,\mathbf{p}\right) =e\hbar \langle v\tau \mathbf{p}%
|i\partial _{\mathbf{p}}|c\tau \mathbf{p}\rangle $. To separate the
topological part of interaction we will consider the electron-doped system
neglecting interband transitions. The conditions which justify this will be
presented further. Thus, we can consider only conduction band dynamics with
the total Hamiltonian%
\begin{equation*}
\hat{H}^{\prime }=\sum_{\tau ,\mathbf{p}}\mathcal{E}_{c\tau }\left( \mathbf{p%
}\right) \hat{e}_{c\tau \mathbf{p}}^{\dagger }\hat{e}_{c\tau \mathbf{p}}+e%
\mathbf{E}(t)i\hbar \sum\limits_{\tau ,\mathbf{p,p}^{\prime }}\delta _{%
\mathbf{p}^{\prime }\mathbf{p}}\partial _{\mathbf{p}^{\prime }}\hat{e}%
_{c\tau \mathbf{p}}^{\dagger }\hat{e}_{c\tau \mathbf{p}^{\prime }}
\end{equation*}%
\begin{equation}
+e\mathbf{E}(t)\sum\limits_{\tau ,\mathbf{p}}\mathbf{A}_{\tau }\left( 
\mathbf{p}\right) \widehat{e}_{c\tau \mathbf{p}}^{+}\widehat{e}_{c\tau 
\mathbf{p}}.  \label{sbH}
\end{equation}%
The response of the system to electromagnetic wave is determined by the
intraband current density $\mathbf{j}=-e\left\langle d\widehat{\mathbf{r}}%
/dt\right\rangle $, which from Heisenberg equation can be written as 
\begin{equation}
\mathbf{j}=-e\frac{1}{i\hbar }\left\langle \left[ \widehat{\mathbf{r}}_{i}+%
\widehat{\mathbf{r}}_{\mathrm{B}},\widehat{H}^{\prime }\right] \right\rangle
.  \label{intraj}
\end{equation}%
Taking into account fermion anticommutator rules, with the help of Eqs. (\ref%
{ri}), (\ref{rb}), and (\ref{sbH}) from Eq. (\ref{intraj}) one can obtain
the intraband surface current 
\begin{equation}
\mathbf{j}\left( t\right) =-eg\sum_{\mathbf{p}\tau }\left( \mathbf{V}_{\tau
}\left( \mathbf{p}\right) +\frac{e}{\hbar }\left[ \mathbf{E}(t)\times 
\mathbf{B}_{\tau }\left( \mathbf{p}\right) \right] \right) \mathcal{N}%
_{c\tau }\left( \mathbf{p},t\right) ,  \label{Jtot}
\end{equation}%
where $\mathbf{V}_{\tau }\left( \mathbf{p}\right) =\partial \mathcal{E}%
_{c\tau }\left( \mathbf{p}\right) /\partial \mathbf{p}$ is the velocity of
band, 
\begin{equation}
\mathbf{B}_{\tau }\left( \mathbf{p}\right) =\hbar \mathrm{rot}_{\mathbf{p}%
}\left( \mathbf{A}_{\tau }\left( \mathbf{p}\right) \right)   \label{BB}
\end{equation}%
is the Berry curvature, $\mathcal{N}_{c\tau }\left( \mathbf{p},t\right)
=\langle \hat{e}_{c\tau \mathbf{p}}^{\dagger }\left( t\right) \hat{e}_{c\tau 
\mathbf{p}}\left( t\right) \rangle $ is the distribution function of
electrons, and $g$ is the degeneracy factor. The second term in Eq. (\ref%
{Jtot}) is an \textquotedblleft anomalous current\textquotedblright\ \cite%
{Kar-Lut}\ represented in terms of the Berry curvature. From Heisenberg
equation $i\hbar \partial \hat{e}_{c\tau \mathbf{p}}^{\dagger }\hat{e}%
_{c\tau \mathbf{p}}/\partial t=\left[ \hat{e}_{c\tau \mathbf{p}}^{\dagger }%
\hat{e}_{c\tau \mathbf{p}},\widehat{H}^{\prime }\right] $ one can also
obtain evolutionary equations for $\mathcal{N}_{c\tau }\left( \mathbf{p}%
,t\right) $. In addition we will assume that the system relaxes at a rate $%
\gamma $ to the equilibrium $\mathcal{N}_{c\tau }^{(0)}\left( \mathbf{p}%
\right) $ distribution. Thus, we obtain the Boltzmann equation for the
distribution of electrons%
\begin{equation}
\frac{\partial \mathcal{N}_{c\tau }\left( \mathbf{p},t\right) }{\partial t}-e%
\mathbf{E}\left( t\right) \frac{\partial \mathcal{N}_{c\tau }\left( \mathbf{p%
},t\right) }{\partial \mathbf{p}}=-\gamma \left( \mathcal{N}_{c\tau }\left( 
\mathbf{p},t\right) -\mathcal{N}_{c\tau }^{(0)}\left( \mathbf{p}\right)
\right) .  \label{deq}
\end{equation}%
We construct $\mathcal{N}_{c\tau }^{(0)}$ from the filling of electron
states according to the Fermi--Dirac-distribution $f_{FD}$: 
\begin{equation}
\mathcal{N}_{c\tau }^{(0)}\left( \mathbf{p}\right) =f_{FD}\left( \mathcal{E}%
_{c\tau }\left( \mathbf{p}\right) \right) =\frac{1}{1+\exp \left( \frac{%
\mathcal{E}_{c\tau }\left( \mathbf{p}\right) -\varepsilon _{F}}{T}\right) }.
\label{fd}
\end{equation}%
Here $\varepsilon _{F}$ is the Fermi energy and $T$ is the temperature in
energy units. Note that this relaxation approximation provides an accurate
description for optical field components oscillating at frequencies $\omega
>>\gamma $.

The differential equation (\ref{deq}) can be solved exactly by the method of
characteristics, which yields\cite{Peres} 
\begin{equation}
\mathcal{N}_{c\tau }\left( \mathbf{p},t\right) =\gamma
\int\nolimits_{-\infty }^{t}dt^{\prime }e^{-\gamma \left( t-t^{\prime
}\right) }\mathcal{N}_{c\tau }^{(0)}(\mathbf{p+p}_{E}\left( t,t^{\prime
}\right) ),  \label{msol}
\end{equation}%
where 
\begin{equation}
\mathbf{p}_{E}\left( t,t^{\prime }\right) =e\int\nolimits_{t^{\prime }}^{t}%
\mathbf{E}\left( t^{\prime \prime }\right) dt^{\prime \prime }.  \label{gmom}
\end{equation}%
is the momentum given by the wave field.

We will consider the case of a low frequency driving wave $2\varepsilon
_{F}/\hbar \omega >>1$. It is clear, that a strong field will induce
multiphoton ($n_{0}>2\varepsilon _{F}/\hbar \omega >>1$) and/or tunneling
transitions from the valence to the conduction band. To neglect the
interband transitions one should restrict the field strength. When a gapped
sample is exposed to an intense laser field the interband transition
mechanisms are distinguished by the Keldysh\cite{Keldysh64} parameter $%
\gamma _{K}=\delta _{t}\omega $, where $\delta _{t}$\ is the tunneling time.%
\textrm{\ }In considered case\ the tunneling time is determined by the mean
time of the electron passing through a barrier of width $2\varepsilon
_{F}/eE_{0}$\ with the velocity $\mathrm{v}_{x}$. Thus, the Keldysh
parameter will be\textrm{\ }%
\begin{equation}
\gamma _{K}=\frac{2\varepsilon _{F}\omega }{eE_{0}\mathrm{v}_{x}}.
\label{kel}
\end{equation}%
At the satisfaction of the condition $2\varepsilon _{F}/\hbar \omega >>1$\
in the strong fields when $\gamma _{K}<<1$, the interband transitions take
place via the tunneling. Thus, we will consider the so-called nonadiabatic
regime $\gamma _{K}\gtrsim 1$\ when tunneling is suppressed and the
probabilities of multiphoton transitions are negligible since $n_{0}>>1$.
This is a quasiclassical regime when the wave-particle interaction can be
characterized by the amplitude of the energy $eE_{0}\mathrm{v}_{x}/\omega $\
of an electron oscillatory motion in the wavefield. The condition $\gamma
_{K}>1$\ can be written as 
\begin{equation}
\frac{eE_{0}\mathrm{v}_{x}}{\omega }<2\varepsilon _{F}.  \label{inter}
\end{equation}%
This means that the wavefield can not provide a sufficient energy for the
creation of an electron-hole pair.

With the help of the solution (\ref{msol}) the topological part of the
induced current can be represented as%
\begin{equation*}
\mathbf{j}_{\mathrm{T}}\left( t\right) =-\frac{e^{2}\gamma g}{4\pi ^{2}\hbar
^{3}}\sum_{\tau }\int\nolimits_{-\infty }^{t}dt^{\prime }\int d\mathbf{p}%
\left( \mathbf{E}(t)\times \mathbf{B}_{\tau }\left( \mathbf{p}\right) \right)
\end{equation*}%
\begin{equation}
\times e^{-\gamma \left( t-t^{\prime }\right) }\mathcal{N}_{c\tau }^{(0)}(%
\mathbf{p+\mathbf{p}}_{E}\left( t,t^{\prime }\right) \mathbf{).}
\label{main1}
\end{equation}%
Taking into account the time reversal symmetry: $\mathcal{N}_{c\tau
}^{(0)}\left( \mathbf{p}\right) =\mathcal{N}_{c-\tau }^{(0)}\left( -\mathbf{p%
}\right) $ and $\mathbf{B}_{\tau }\left( \mathbf{p}\right) =-\mathbf{B}%
_{-\tau }\left( -\mathbf{p}\right) $ the $\mathbf{j}_{\mathrm{T}}$ can be
expressed as 
\begin{equation*}
\mathbf{j}_{\mathrm{T}}\left( t\right) =-\frac{e^{2}\gamma g}{4\pi ^{2}\hbar
^{2}}\int d\mathbf{p}\int\nolimits_{-\infty }^{t}dt^{\prime }e^{-\gamma
\left( t-t^{\prime }\right) }\mathcal{N}_{c1}^{(0)}(\mathbf{p})
\end{equation*}%
\begin{equation}
\times \left[ \mathbf{E}(t)\times \left( \mathbf{B}_{1}\left( \mathbf{p-p}%
_{E}\left( t,t^{\prime }\right) \right) -\mathbf{B}_{1}\left( \mathbf{p+p}%
_{E}\left( t,t^{\prime }\right) \right) \right) \right] .  \label{Main}
\end{equation}%
This result provides a near-analytical expression for the topological
current including contributions from all orders in the field. When $\mathbf{E%
}\left( t\right) =0$, $\mathbf{j}_{\mathrm{T}}=0$. Since $\mathbf{B}\left( 
\mathbf{p}\right) =\mathbf{B}\left( -\mathbf{p}\right) $, when a tilt is
absent $\alpha =0$, then $\mathcal{N}_{c1}^{(0)}(\mathbf{p})=\mathcal{N}%
_{c1}^{(0)}\left( -\mathbf{p}\right) $, and $\mathbf{j}_{\mathrm{T}}=0$. The
net topological current is vanishing also when the field is perpendicular to
tilt direction. From the dependence on the field $\mathbf{j}_{\mathrm{T}%
}\left( \mathbf{E}\right) =\mathbf{j}_{\mathrm{T}}\left( -\mathbf{E}\right) $
it is apparent that the topological current contains only even orders of
nonlinear response and is directed perpendicular to the pump field. As is
seen the topological current strongly depends on tilt parameter $\alpha $.
The leading order $\sim \alpha $\ of which can be obtained from Eq. (\ref%
{Main}) by expanding Fermi--Dirac-distribution function:\textrm{\ }%
\begin{equation*}
\mathbf{j}_{\mathrm{T}}\left( t\right) \simeq \alpha \frac{e^{2}\gamma g}{%
2\pi ^{2}\hbar ^{2}}\int d\mathbf{p}\int\nolimits_{-\infty }^{t}dt^{\prime
}e^{-\gamma \left( t-t^{\prime }\right) }p_{x}\frac{\partial f_{FD}\left( 
\mathcal{E}\right) }{\partial \mathcal{E}}
\end{equation*}%
\begin{equation}
\times \left[ \mathbf{E}(t)\times \mathbf{B}_{1}\left( \mathbf{p+p}%
_{E}\left( t,t^{\prime }\right) \right) \right] .  \label{alfa}
\end{equation}

The regular part of the current can be represented as%
\begin{equation*}
\mathbf{j}_{\mathrm{r}}\left( t\right) =-\frac{e\gamma g}{4\pi ^{2}\hbar ^{2}%
}\sum_{\tau }\int\nolimits_{-\infty }^{t}dt^{\prime }e^{-\gamma \left(
t-t^{\prime }\right) }
\end{equation*}%
\begin{equation}
\times \int d\mathbf{pV}_{\tau }\left( \mathbf{p-p}_{E}\left( t,t^{\prime
}\right) \right) \mathcal{N}_{c\tau }^{(0)}\left( \mathbf{p}\right) .
\label{reg1}
\end{equation}%
Taking into account the time reversal symmetry $\mathbf{V}_{1}\left( \mathbf{%
p}\right) =\mathbf{V}_{-1}\left( -\mathbf{p}\right) $ it can be expressed as 
\begin{equation*}
\mathbf{j}_{\mathrm{r}}\left( t\right) =-\frac{e\gamma g}{4\pi ^{2}\hbar ^{2}%
}\int\nolimits_{-\infty }^{t}dt^{\prime }e^{-\gamma \left( t-t^{\prime
}\right) }\int d\mathbf{p}
\end{equation*}%
\begin{equation}
\times \left[ \mathbf{V}_{1}\left( \mathbf{p-p}_{E}\left( t,t^{\prime
}\right) \right) -\mathbf{V}_{1}\left( \mathbf{p+p}_{E}\left( t,t^{\prime
}\right) \right) \right] \mathcal{N}_{c1}^{(0)}(\mathbf{p}).  \label{reg}
\end{equation}%
From Eq. (\ref{reg}) it is clear that $\mathbf{j}_{\mathrm{r}}\left( \mathbf{%
E}\right) =-\mathbf{j}_{\mathrm{r}}\left( -\mathbf{E}\right) $. The regular
current contains only odd orders of nonlinear response and is directed along
the pump field. When a tilt is absent $\alpha =0$, then $\mathcal{N}%
_{c1}^{(0)}(\mathbf{p})=\mathcal{N}_{c1}^{(0)}\left( -\mathbf{p}\right) $
and the regular current can be written as: 
\begin{equation}
\mathbf{j}_{\mathrm{r}}\left( t\right) =-\frac{e\gamma g}{2\pi ^{2}\hbar ^{2}%
}\int\nolimits_{-\infty }^{t}dt^{\prime }e^{-\gamma \left( t-t^{\prime
}\right) }\int d\mathbf{pV}_{1}\left( \mathbf{p-p}_{E}\left( t,t^{\prime
}\right) \right) \mathcal{N}_{c1}^{(0)}(\mathbf{p}).  \label{reg2}
\end{equation}%
The last formula represents total interband current when three-fold ($C_{3}$%
) symmetry is preserved. On the base of Eq. (\ref{reg2}) in the
nonperturbative regime with corresponding velocity $\mathbf{V}_{1}\left( 
\mathbf{p}\right) $ dependence on $\mathbf{p}$ the HHG processes have been
investigated in graphene\cite{Mikhailov} and in monolayer black phosphorus.%
\cite{bf}

To proceed it is also useful to present multipole expansion of the
topological current (\ref{Main}). For concreteness we will consider
x-polarized wave $\widehat{\mathbf{e}}=\widehat{\mathbf{x}}$\ and utilize
the Taylor expansion of the Berry curvature in Eq. (\ref{Main}) 
\begin{equation*}
B_{1z}\left( p_{x}+p_{E},p_{y}\right) =\sum\limits_{n=0}^{\infty }\frac{%
\partial ^{n}B_{1z}\left( p_{x},p_{y}\right) }{\partial p_{x}^{n}}\frac{%
p_{E}^{n}}{n!}.
\end{equation*}%
Taking into account that only odd orders of $p_{E}^{n}$\ make contribution
we obtain multipole expansion of the topological current%
\begin{equation}
j_{y\mathrm{T}}\left( t\right) =-2e^{2}\gamma \sum\limits_{n=1}^{\infty }%
\frac{D_{x}^{\left( 2n-1\right) }}{\left( 2n-1\right) !}\int\nolimits_{-%
\infty }^{t}dt^{\prime }e^{-\gamma \left( t-t^{\prime }\right)
}E_{x}(t)p_{E}^{2n-1}\left( t,t^{\prime }\right) ,  \label{multi}
\end{equation}%
where\textrm{\ }%
\begin{equation}
D_{x}^{\left( 2n-1\right) }=g\int \frac{d\mathbf{p}}{\left( 2\pi \hbar
\right) ^{2}}\mathcal{N}_{c1}^{(0)}(\mathbf{p})\frac{\partial
^{2n-1}B_{1z}\left( p_{x},p_{y}\right) }{\partial p_{x}^{2n-1}}
\label{multipole}
\end{equation}%
is the $\left( 2n-1\right) $-th moment of the Berry curvature in the
momentum space. As is expected, in the leading order topological current is
defined by the dipole moment of the Berry curvature. This corresponds to
second order nonlinear Hall current derived by Sodemann and Fu.\cite{Sod-Fu}
The next nonvanishing moment is the octupole moment. From Eq. (\ref{multi})
it is easy to extract low-field perturbative limit of the topological
current when $\left\vert p_{E}\right\vert <<p_{F}$\ ($p_{F}$\ is the Fermi
momentum). At $\gamma <<\omega $\ and for continuos wave we obtain
perturbative limit of the topological current%
\begin{equation*}
j_{y\mathrm{T}}\left( t\right) =\sum\limits_{n=1}^{\infty }j_{y}^{(2n\omega
)}\left( t\right) ,
\end{equation*}%
where\textrm{\ }%
\begin{equation}
j_{y}^{(2n\omega )}\left( t\right) =\frac{e^{2n+1}D_{x}^{\left( 2n-1\right)
}E_{0}^{2n}}{\omega ^{2n-1}2^{2n-2}\left( 2n-1\right) !}\sin 2n\omega t.
\label{2np}
\end{equation}%
As is seen, in the perturbative limit the $\left( 2n\right) $-th harmonic is
defined by the $\left( 2n-1\right) $-th moment of the Berry curvature. Note
that for strong fields the current corresponding to $\left( 2n\right) $-th
harmonic may have sizable contribution from higher moments.

\section{Results}

We further examine the nonlinear response of a nanostructure considering the
generation of harmonics at the multiphoton excitation. We will consider
low-temperature limit $T<<\varepsilon _{F}$. In this case the main
contribution in the integrals determining topological (\ref{Main}) and
regular (\ref{reg}) currents is made by the Fermi surface. We aim to keep
the consideration rather generic introducing a limited number of
dimensionless parameters. From Eqs. (\ref{Bcon}) and (\ref{BB}) one can
calculate the Berry curvature 
\begin{equation}
\mathbf{B}_{\tau }\left( \mathbf{p}\right) =-\frac{\Delta }{2}\tau \widehat{%
\mathbf{z}}\hbar ^{2}\frac{\mathrm{v}_{x}\mathrm{v}_{y}}{\left( \Delta ^{2}+%
\mathrm{v}_{x}^{2}p_{x}^{2}+\mathrm{v}_{y}^{2}p_{y}^{2}\right) ^{3/2}}.
\label{Bc}
\end{equation}%
The band velocity $V_{x\tau }\left( \mathbf{p}\right) =\partial \mathcal{E}%
_{c\tau }\left( \mathbf{p}\right) /\partial p_{x}$ can also be calculated,
which gives 
\begin{equation}
V_{x\tau }\left( \mathbf{p}\right) =\alpha \tau +\frac{\mathrm{v}%
_{x}^{2}p_{x}}{\left( \Delta ^{2}+\mathrm{v}_{x}^{2}p_{x}^{2}+\mathrm{v}%
_{y}^{2}p_{y}^{2}\right) ^{1/2}}.  \label{Vx}
\end{equation}%
As is seen from Eq. (\ref{Bc}) the tilting effect does not change the Berry
curvature of the system but it is crucial to get a corresponding
nonvanishing net topological current. The latter is determined by the
odd-order moments of the Berry curvature in the momentum space (\ref%
{multipole}). That is, one needs shear or warping of the Fermi surface: $%
N_{c\tau }^{(0)}(\mathbf{p})\neq N_{c\tau }^{(0)}\left( -\mathbf{p}\right) $%
. In considered case this is satisfied, since the Fermi surface is%
\begin{equation}
\alpha \tau p_{x}+\sqrt{\Delta ^{2}+\mathrm{v}_{x}^{2}p_{x}^{2}+\mathrm{v}%
_{y}^{2}p_{y}^{2}}=\varepsilon _{F}.  \label{Fs}
\end{equation}%
Making transformation $p_{y}\rightarrow \mathrm{v}_{x}/\mathrm{v}_{y}p_{y}$
one can see that the Fermi surface is an ellipse in the new coordinates (we
assume that $\varepsilon _{F}>\Delta $). At $\alpha =0$ Fermi surface is a
circle with Fermi momentum $p_{F}=\sqrt{\varepsilon _{F}^{2}-\Delta ^{2}}/%
\mathrm{v}_{x}$. In Eq. (\ref{Main}) for integration we normalize $\mathbf{p}
$ by the Fermi momentum and write $\overline{p}_{x}=p_{x}/p_{F}$, $\overline{%
p}_{y}\rightarrow \mathrm{v}_{y}/\mathrm{v}_{x}p_{y}/p_{F}$, $\overline{p}%
_{E}=p_{xE}/p_{F}$, and $\overline{\Delta }=\Delta /\left( \mathrm{v}%
_{x}p_{F}\right) $ so that%
\begin{equation*}
j_{y\mathrm{T}}\left( t\right) =\frac{e^{2}g}{4\pi ^{2}\hbar }E_{x}(t)\frac{%
\overline{\Delta }}{2}\gamma \int\nolimits_{-\infty }^{t}dt^{\prime
}e^{-\gamma \left( t-t^{\prime }\right) }\int d\overline{p}_{x}d\overline{p}%
_{y}
\end{equation*}%
\begin{equation}
\times \frac{\mathcal{N}_{c1}^{(0)}(\overline{\mathbf{p}})-\mathcal{N}%
_{c-1}^{(0)}(\overline{\mathbf{p}})}{\left( \overline{\Delta }^{2}+\left( 
\overline{p}_{x}+\overline{p}_{E}\left( t,t^{\prime }\right) \right) ^{2}+%
\overline{p}_{y}^{2}\right) ^{3/2}},  \label{jyt}
\end{equation}%
where 
\begin{equation}
\overline{p}_{E}\left( t,t^{\prime }\right) =\frac{F_{0}}{\sqrt{1-\frac{%
\Delta ^{2}}{\varepsilon _{F}^{2}}}}\int\nolimits_{t^{\prime }}^{t}f\left(
t^{\prime \prime }\right) \cos \left( \omega t^{\prime \prime }\right)
d\left( \omega t^{\prime \prime }\right) .  \label{pe}
\end{equation}%
The dimensionless interaction parameter is defined as $F_{0}=eE_{0}\mathrm{v}%
_{x}/\left( \omega \varepsilon _{F}\right) $. $\ $The interaction parameter $%
F_{0}$ is subject to the constraint (\ref{inter}), which yields $F_{0}<2$.
Overall the topological current is defined by the following parameters: $%
F_{0}$, $\Delta /\varepsilon _{F}$, $\gamma /\omega $, $\alpha /\mathrm{v}%
_{x}$, and $\Delta /\left( \hbar \omega \right) $. A similar equation can be
obtained for the regular current. In general, for strong pump waves $%
F_{0}\sim 1$ the integration in Eq. (\ref{jyt}) can not be made analytically
and one should integrate it numerically. Thus, making integration in Eq. (%
\ref{jyt}), one can calculate the harmonic radiation spectrum with the help
of a Fourier transform $j_{y\mathrm{T}}\left( \omega \right) $ of the
function $j_{y\mathrm{T}}\left( t\right) $. Note that for a sufficiently
large 2D sample the generated field will be $-4\pi j_{y\mathrm{T}}\left(
t\right) /c$. Hence, we will characterize the emission strength of the $s$th
harmonic by the dimensionless parameter $\eta _{y}\left( s\right) =4\pi
\left\vert j_{y\mathrm{T}}\left( s\omega \right) \right\vert /cE_{0}$.
Similarly, we will characterize harmonic radiation due to the regular part
of the current $\eta _{x}\left( s\right) $.

We start by investigating the field dependence of the harmonics radiation
due to Berry curvature of the bands that induce anomalous current
perpendicular to the applied pump field. To ensure a dominating intraband
response we need low photon energies $\hbar \omega <<\varepsilon _{F}$ and a
sufficient number of free carriers. We will consider two types of systems
with relativistic energy bands: semiconductors $\varepsilon _{F}\gtrsim
\Delta $ such as transition metal dichalcogenides, or graphenlike semimetals
with $\varepsilon _{F}>>\Delta $, such as gaped graphene, topological
crystalline insulator, silicene, germanene, where a small gap is opened by
the spatial inversion symmetry breaking. A damping rate $\gamma /\omega
=0.05 $ and temperature $T/\left( \mathrm{v}_{x}p_{F}\right) =0.01$ will be
assumed in all plots below. 
\begin{figure}[tbp]
\includegraphics[width=.45\textwidth]{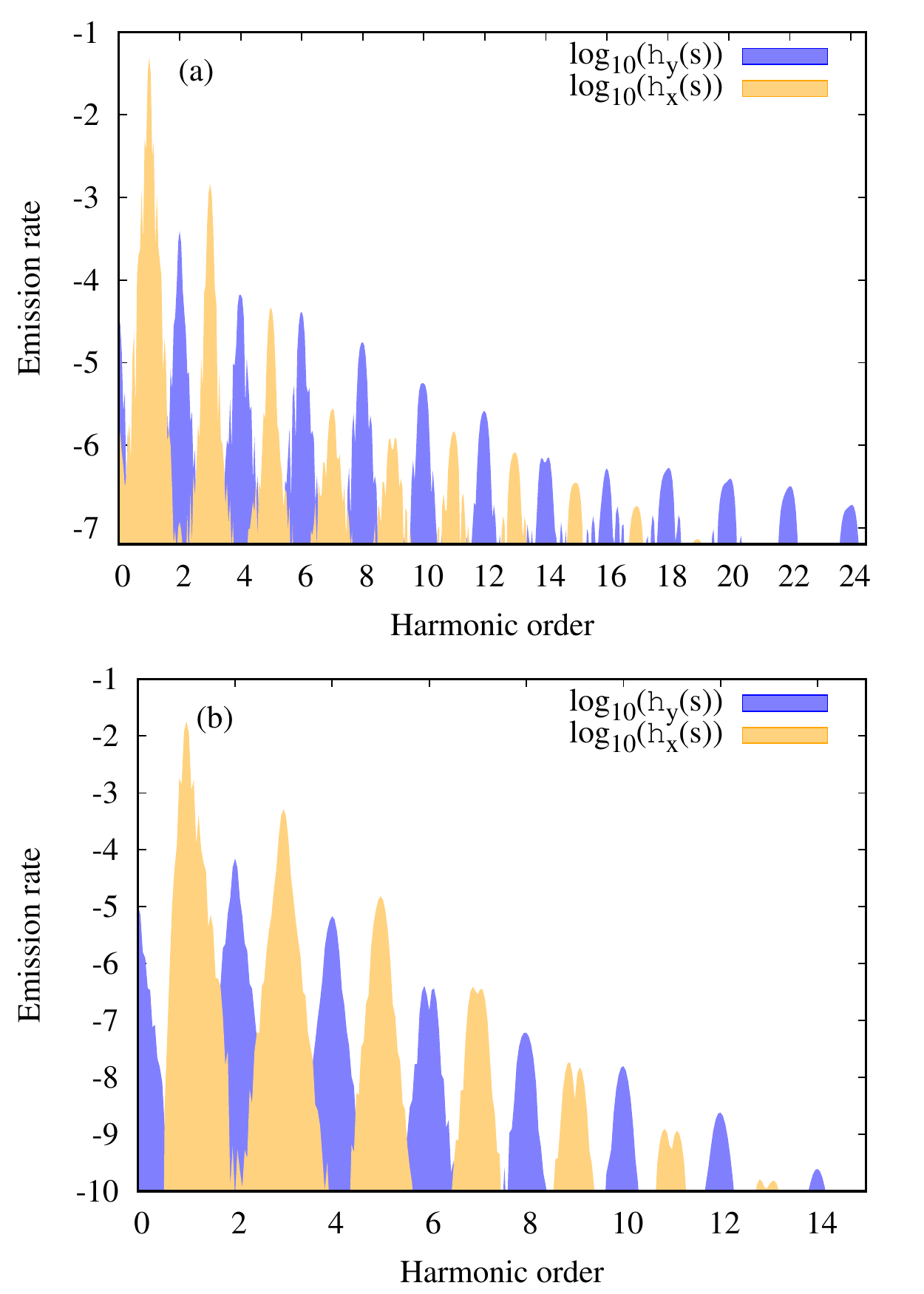}
\caption{Polarization resolved radiation spectrum via the logarithm of the
normalized field strengths $\protect\eta _{x,y}\left( s\right) $ (in
arbitrary units): (a) for $\protect\chi =1$, $\Delta =\hbar \protect\omega $%
, $\protect\varepsilon _{F}=10\Delta $, $\protect\alpha =0.2\mathrm{v}_{x}$;
(b) for $F_{0}=1$, $\Delta =10\hbar \protect\omega $, $\protect\varepsilon %
_{F}=1.2\Delta $ and $\protect\alpha =0.2\mathrm{v}_{x}$.}
\end{figure}
\begin{figure}[tbp]
\includegraphics[width=.45\textwidth]{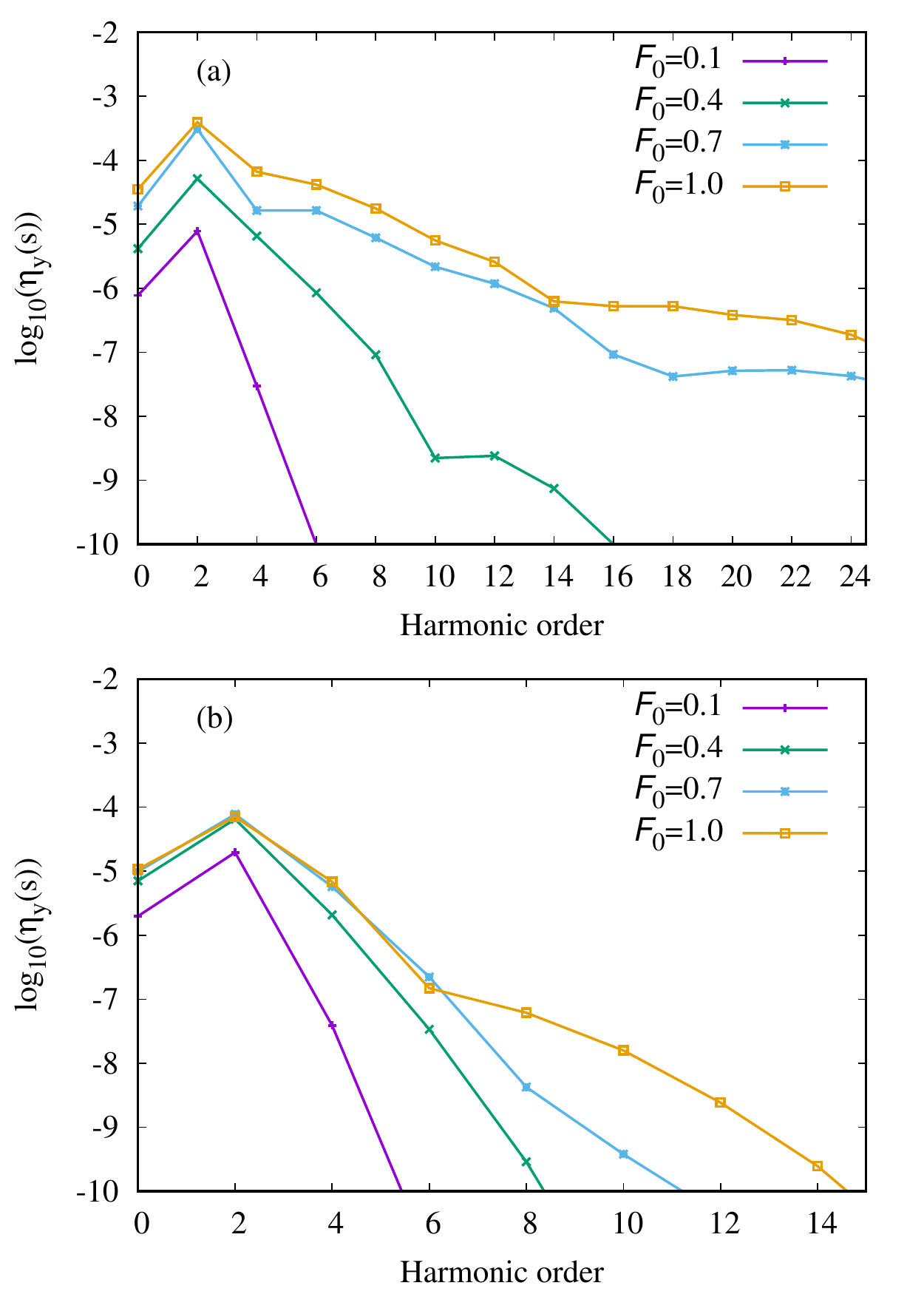}
\caption{The Fourier content of the topological current. The radiation
spectrum induced by Berry curvature for a range of field strengths: (a) for $%
\Delta =\hbar \protect\omega $, $\protect\varepsilon _{F}=10\Delta $, and $%
\protect\alpha =0.2\mathrm{v}_{x}$; (b) for $\Delta =10\hbar \protect\omega $%
, $\protect\varepsilon _{F}=1.2\Delta $, and $\protect\alpha =0.2\mathrm{v}%
_{x}$. }
\end{figure}
In Fig. 1 the polarization-resolved radiation spectrum via the logarithm of
the normalized field strengths are displayed $\eta _{x,y}\left( s\right) $
at strong pump wave $F_{0}=1$ for semiconductors, and for semimetals. As is
seen from this figure, in both cases the higher harmonics due to topological
current are dominated over regular ones. This tendency is more strict in the
case of semimetals.

The Fourier content of the topological current for a range of field
strengths is shown in Fig. 2. We have considered two cases in Fig. 2: $%
\varepsilon _{F}=10\Delta $ (upper panel) and $\varepsilon _{F}=1.2\Delta $
(lower panel). At low fields, the HHG emission rates decrease rapidly with
the order of harmonic. While a much more gradual decrease is observed for
semimetals at large $F_{0}$ $\sim 1.0$. This can be explained as follow. It
is well known that the Fourier transform is notably sensitive to
singularities. The Fourier image of a real analytic function is
exponentially decaying at high frequencies.\cite{Reed} However, if there is
a discontinuity, its Fourier image decays according to a power law. The
exponent is determined by the type of singularity. As is seen from Eq. (\ref%
{jyt}), for semimetals when $\overline{\Delta }\rightarrow 0$ in the
denominator we have singularity which provides plateau for HHG emission
rate. For regular current this is well seen for graphene, when $\Delta =0$.%
\cite{Mikhailov} To emphasize this finding, in Fig. 3 we plot the Fourier
content of radiation induced by the Berry curvature for different Fermi
energies at fixed pump wave intensity. As is seen from this figure, with the
increasing Fermi energy we have a more gradual decrease in emission rate
depending on the harmonic order. In Fig. 3 we have included the case $%
\varepsilon _{F}=2\Delta $\ even though this is clearly outside the
intraband regime. However, we want to demonstrate a plateau formation as the
gap narrows.

\begin{figure}[tbp]
\includegraphics[width=.45\textwidth]{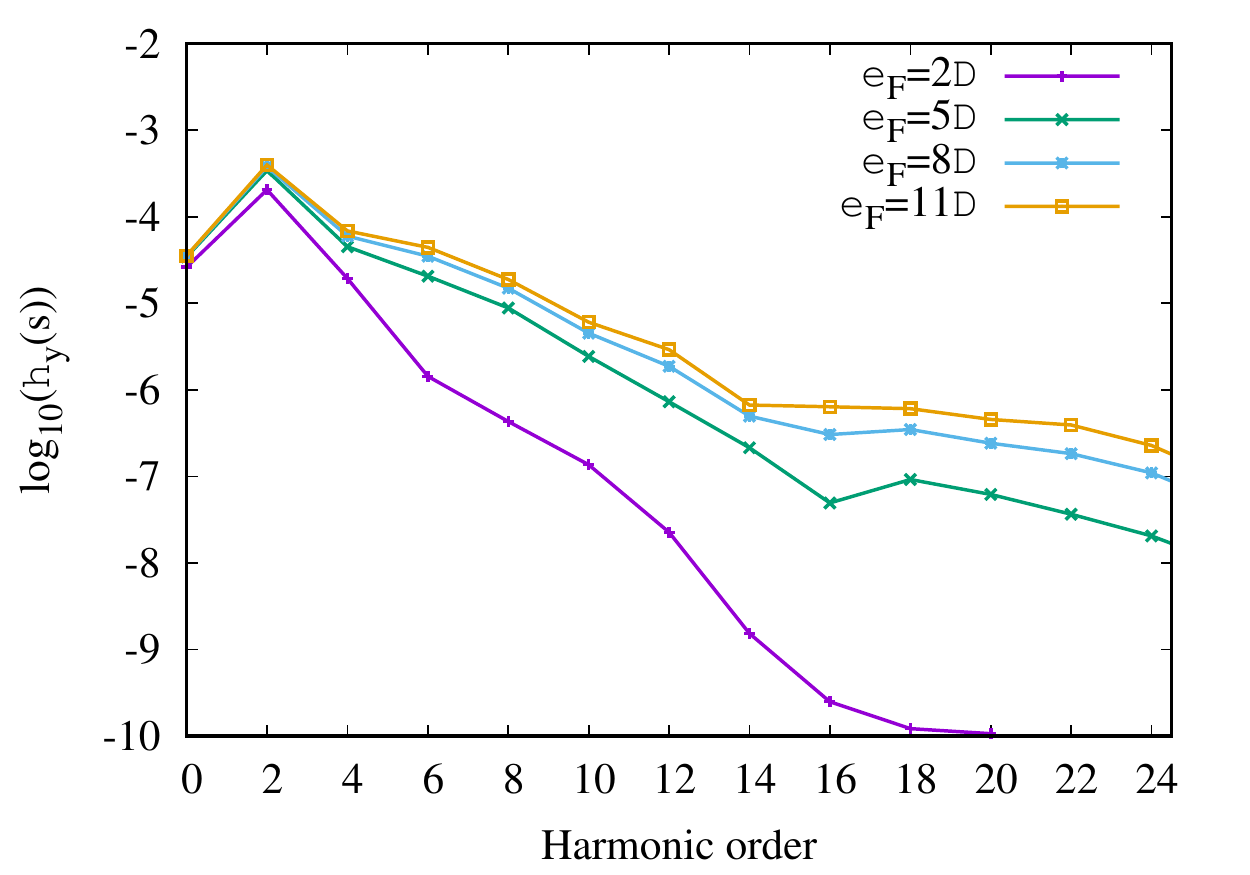}
\caption{The Fourier content of radiation, induced by the Berry curvature
for different Fermi energies at $F_{0}=1$, $\Delta =\hbar \protect\omega $,
and $\protect\alpha =0.2\mathrm{v}_{x}$.}
\end{figure}

The net topological current strongly depends on the tilt parameter $\alpha $%
. To show this dependence, in Fig. 4 we plot Fourier content of
radiation-induced by the Berry curvature at the fixed pump wave intensity $%
F_{0}=1$\ for various values of $\alpha $.\ For both cases, at moderate
harmonics the emission rates $\sim \alpha $\ in agreement with Eq. (\ref%
{alfa}). However, at high orders of harmonics this tendency is not
preserved. This can be explained from the multipole expansion of the
topological current Eq. (\ref{multi}). As is clear from Eq. (\ref{multi}),
since $p_{E}\left( t,t^{\prime }\right) \sim \left( \sin \omega t-\sin
\omega t^{\prime }\right) $ the current corresponding to ($2s$)-th harmonic
besides $D_{x}^{\left( 2s-1\right) }$\ have contribution from higher moments 
$D_{x}^{\left( 2n-1\right) }$\ ($n>s$) of the Berry curvature. The higher
moments are more sensitive to Fermi surface deformation which takes place at
the change of the tilt parameter $\alpha $.\ For strong fields, the higher
moments of the Berry curvature have sizable contributions that may sum up as
constructively as well as destructively. The latter explains the behavior of
HHG spectra in Fig. 4 at high orders of harmonics. 
\begin{figure}[tbp]
\includegraphics[width=.45\textwidth]{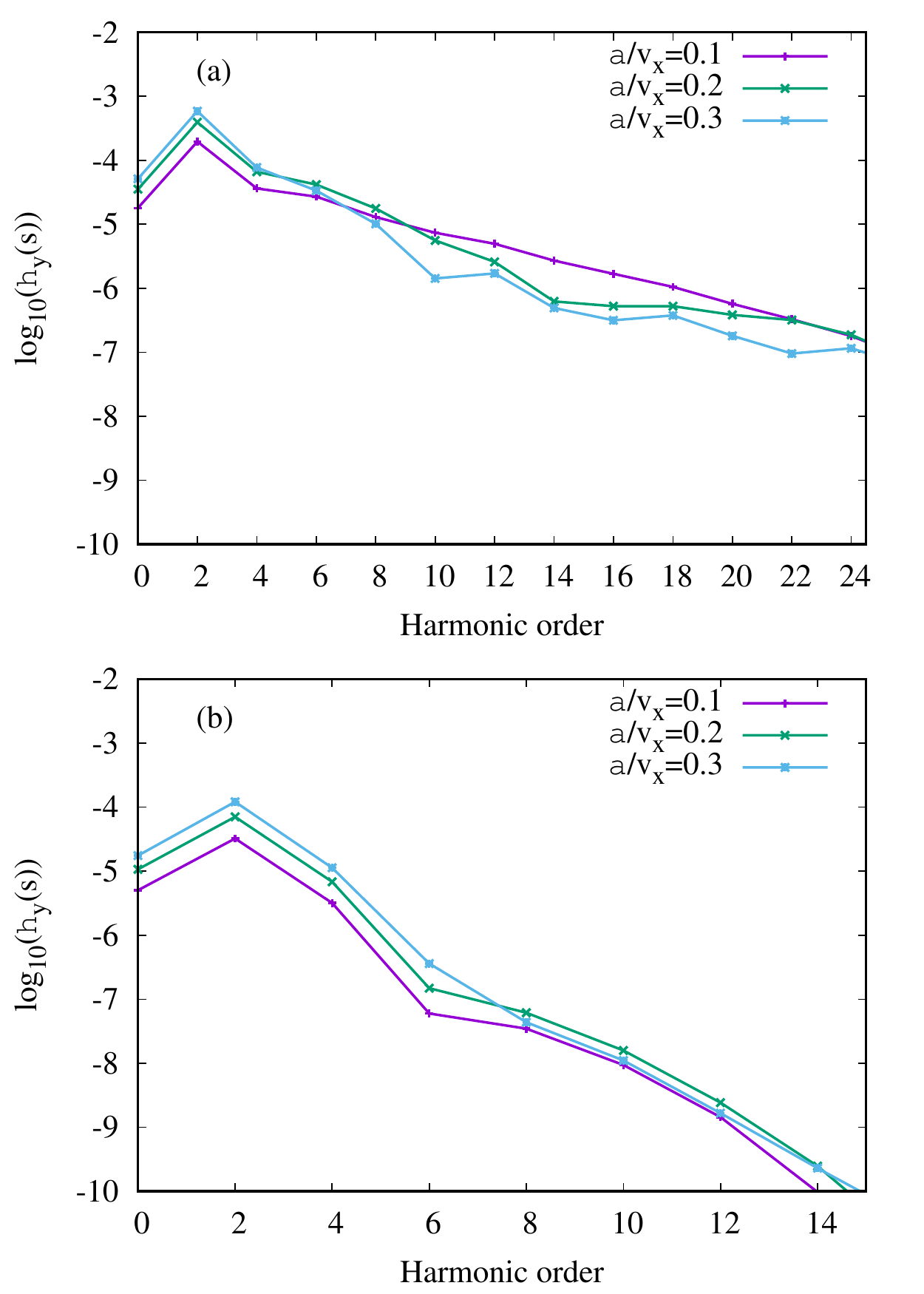}
\caption{The Fourier content of radiation, induced by the Berry curvature at
fixed pump wave normalized field strength $F_{0}=1$ for various values of
the tilt parameter $\protect\alpha $: (a) for $\Delta =\hbar \protect\omega $
and $\protect\varepsilon _{F}=10\Delta $; (b) for $\Delta =10\hbar \protect%
\omega $ and $\protect\varepsilon _{F}=1.2\Delta $.}
\end{figure}

It is also of interest to more accurately determine the field strength
required to achieve a nonperturbative regime of HHG. For this propose in
Fig. 5 we display field dependence of scaled nonlinear conductances $\sigma
_{y}\left( s\right) =\eta _{y}\left( s\right) /F_{0}^{s-1}$\ for the second
and fourth harmonics in two cases.\textrm{\ }In the perturbative limit (\ref%
{2np}), the $s$th harmonic response via the normalized field strengths $\eta
_{y}\left( s\right) $\ varies with the field as $\eta _{y}\left( s\right)
\propto F_{0}^{s-1}$. Hence, the nonlinear conductances $\sigma _{y}\left(
2\right) $\ and $\sigma _{y}\left( 4\right) $\ are field-independent in this
limit.\textrm{\ }The perturbative results are indicated by the solid lines
in Fig. 5. As is seen, the full numerical results agree with these
predictions in the low field limit. For a semimetal case marked deviations
occur at $F_{0}>0.2$ for both second and fourth harmonics. In comparison,
for the semiconductor case the transition is at a smaller field of $%
F_{0}>0.1 $. Thus, the threshold field for nonperturbative regime is
different for semimetals and for semiconductors, and in addition, the
behavior in the nonperturbative regime of HHG is strictly different. For
semimetals, the emission rates are increased more rapidly with the
increasing pump wave intensity. While for semiconductors the emission rates
are increased slowly, thus nonlinear conductances are decreased with the
increasing pump wave intensity. 
\begin{figure}[tbp]
\includegraphics[width=.45\textwidth]{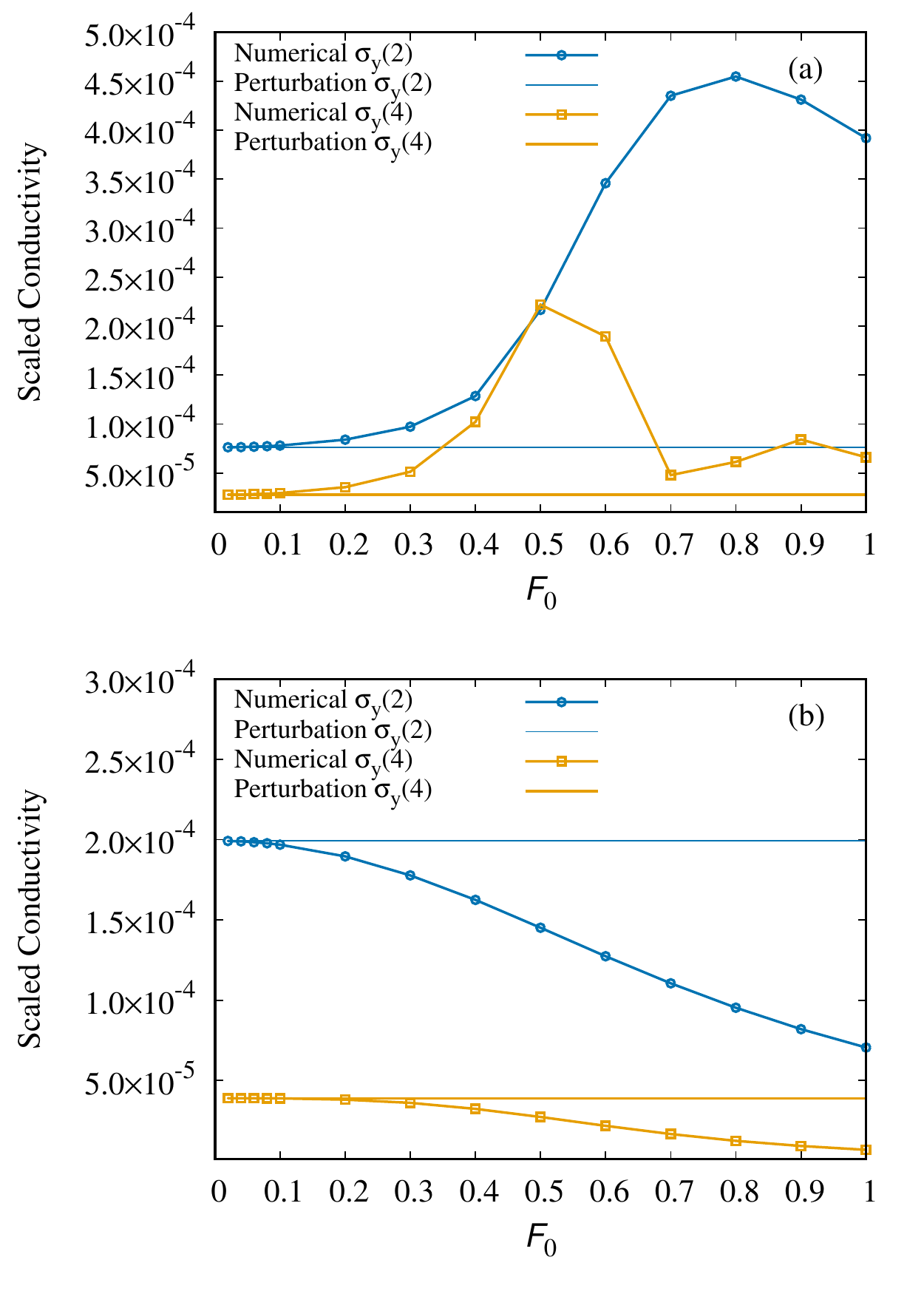}
\caption{Scaled second and fourth order nonlinear conductances $\protect%
\sigma _{y}\left( 2\right) $ and $\protect\sigma _{y}\left( 4\right) $
versus normalized field strength $F_{0}$: (a) for $\Delta =\hbar \protect%
\omega $, $\protect\varepsilon _{F}=10\Delta $, and $\protect\alpha =0.2%
\mathrm{v}_{x}$; (b) for $\Delta =10\hbar \protect\omega $, $\protect%
\varepsilon _{F}=1.2\Delta $, and $\protect\alpha =0.2\mathrm{v}_{x}$.}
\end{figure}
Let us estimate the pump wave parameters for the considered process of HHG
due to topology of bands. The average intensity of the pump wave expressed
by $F_{0}$ can be estimated as%
\begin{equation*}
I=\left( F_{0}\frac{c}{\mathrm{v}_{x}}\frac{\varepsilon _{F}}{\mathrm{eV}}%
\frac{\hbar \omega }{\mathrm{eV}}\right) ^{2}\times 3.\,\allowbreak 4\times
10^{6}\mathrm{\ W\ cm}^{-2},
\end{equation*}%
where $c$ is the light speed in vacuum. The typical values of parameters for
a semimetal case are: $\mathrm{v}_{x}\approx \mathrm{v}_{y}\approx 4\times
10^{7}\mathrm{cm/s}$, $\Delta \approx 10\mathrm{\ meV}$, $\alpha \approx 0.1%
\mathrm{v}_{x}$.\cite{Sod-Fu} Thus, nonperturbative topological HHG will
take place at the photon energy $\hbar \omega \approx 10\mathrm{\ meV}$ and
intensity $7.\,\allowbreak 6\times 10^{4}\mathrm{\ W\ cm}^{-2}$. The
required Fermi energy is\textrm{\ }$\varepsilon _{F}\approx 0.1\mathrm{eV}$%
\textrm{.} For the semiconductor case, we assume transition metal
dichalcogenides with $\Delta \approx 1\mathrm{\ eV}$ and nonperturbative
effects will be essential at $\hbar \omega \approx 0.1\mathrm{eV}$\textrm{, }%
$\varepsilon _{F}\approx 1.2\mathrm{eV,}$ and the wave intensity $%
2.\,\allowbreak 7\times 10^{8}\mathrm{\ W\ cm}^{-2}$.

In the end, let us make a remark about the disorder mediated correction to
the HHG process. As was shown in Refs.\cite{Disorder1,Disorder2,Disorder3}
in addition to the Berry curvature dipole term there exist additional
disorder mediated corrections to the second-order nonlinear Hall tensor that
have the same scaling in impurity scattering rate. For the nonperturbative
regime, this issue demands further consideration. Hence, it is of interest
to clear up disorder-induced contributions to the nonperturbative
topological HHG. The latter requires extensive numerical analysis and will
be the subject of future work.

\section{Conclusion}

In this paper, the nonlinear optical response of electrons in
pseudorelativistic energy bands with nonzero Berry curvature has been
investigated. As a model time-reversal invariant system we have considered
nanostructures where the Dirac cones are tilted. A semianalytical
calculation including all harmonic orders has been presented. We have
studied the harmonic content of the induced topological current and have
shown that HHG induced solely by the Berry curvature of bands is comparable
to regular HHG. At that its polarization is perpendicular to applied pump
wave polarization and consequently to regular HHG ones. We have also studied
the dependence of the nonlinear response on the driving wave and system
parameters for semiconductors and semimetallic cases. It has been shown that
in the graphenlike semimetallic cases the topological HHG spectrum has
plateau character. The field dependence of the HHG reveals threshold field
strengths above which the nonperturbative behavior sets in. Our results
apply to a large number of two-dimensional materials where one can achieve
the minimum symmetry constraints for a non-vanishing topological current.
The corresponding HHG process can thus be used as a way to directly probe
and disclose the geometric properties of energy bands in a large number of
time-reversal invariant materials. In particular, using multipole expansion
of the topological current one can retrieve higher moments of the Berry
curvature from the HHG spectrum. This will be more informative beyond the
Dirac cone approximation and applicable to the full Brillouin zone.

\begin{acknowledgments}
This work was supported by the RA Science Committee, in the frame of
Research Project No. 18T-1C259.
\end{acknowledgments}

\end{document}